\documentclass[aps,pra,reprint,
floatfix,showpacs,groupedaddress,letterpaper]{revtex4-1}

\usepackage{graphicx,subfigure}
\usepackage{dcolumn,setspace}
\usepackage{bm}
\usepackage{hyperref}
\usepackage{amsfonts,amsmath,amssymb}
\usepackage{url,textcase,natbib}
\usepackage{float}
\begin{document}

\title{Robust signals from a quantum-based magnetic compass sensor}

\author{Maria Procopio}
\author{Thorsten Ritz}
\affiliation{Department of Physics and Astronomy, University of California at Irvine, Irvine, CA 92697, USA}

\date{\today}

\begin{abstract}
A quantum-based magnetic compass sensor, mediated through radical pair reactions, has been suggested to underlie the sensory ability of migrating birds to receive directional information from the geomagnetic field. Here we extend the currently available models by considering the effects of slow fluctuations in the nuclear spin environment on the directional signal. We quantitatively evaluate the robustness of signals under fluctuations on a timescale longer than the lifetime of a radical pair, utilizing two models of radical pairs. Our results suggest design principles for building a radical-pair based compass sensor that is both robust and highly directional sensitive.
\end{abstract}

\graphicspath{{./}}



\maketitle
\section{\label{sec:intro}Introduction}
Behavioural experiments suggest that migratory birds possess a magnetic compass sensor able to detect the direction of the geomagnetic field \cite{WWbook,WW2006}. One hypothesis for the basis of this remarkable sensory ability is that the coherent quantum spin dynamics of photoinduced radical pair reactions transduces directional magnetic information from the geomagnetic field into changes of reaction yields \cite{Schulten1978,Ritz2004}, possibly involving the photoreceptor cryptochrome in the birds retina \cite{Ritz2000,Maeda2012,Mouritsen2004}.

The suggested radical-pair based magnetic compass sensor has attracted attention in the field of quantum biology \cite{Nori2012,Ritzquantum} as an example of a biological sensor which might exploit quantum coherences for its biological function. Uncovering the design principles of such a spin-based sensor can pave the way for the design of man-made quantum-based magnetic sensors \cite{Maeda2008}.

Investigations on radical-pair based compass sensors have focussed on identifying the design features optimizing directional sensitivity to weak magnetic fields \citep{Ritz2009,Cai2012,Sun2012}. In addition, a possible role of non-trivial quantum effects such as entanglement \cite{Vedral2011,Hore2012Enta,Pauls2013} or the quantum Zeno effect \cite{Kominis2009} in magnetoreception has been discussed. Finally, the use of quantum control techniques \cite{Cai2010}, gradient fields via magnetic nanostructures \cite{Cai2011} and photo-control techniques \cite{Guerreschi2013} have been explored as means to enhance magneto-sensitivity.

In the radical pair mechanism, after a radical pair is formed in a spin-conserving reaction from a singlet or triplet precursor, the spin state of the radical pair oscillates via quantum coherences determined, in the simplest formulation, by hyperfine interactions only. An external magnetic field of comparable intensity to the geomagnetic field can influence the spin dynamics via weak Zeeman interactions by unlocking some zero-field degenerate coherences, thus affecting the singlet yield \cite{Timmel1998}. Because hyperfine interactions are generally anisotropic, the singlet yield will depend on the intensity as well as on the direction of the external magnetic field \cite{Timmel2001}.
The dependence of the singlet product yield on the direction of the geomagnetic field is the initial signal containing directional information. The steps transducing this signal further are still unclear \cite{Ritz2000,Weaver2000,Ston2012}.

Regardless of the mechanism of signal transduction, the question arises how one should design a radical pair system such as to optimize its directional sensitivity. Besides kinetic considerations, a key factor is the choice of the nuclear spin environment \cite{RodgersHore2009}. Directional sensitivity is optimized for strong internal magnetic field gradients \cite{Cai2012}, which can be realised by the absence of hyperfine interactions on one radical \cite{Ritz2009,Ref_Probe}. More realistic models include dipolar and electron exchange interactions that can affect singlet yields, thereby reducing the directional sensitivity of a radical-pair based compass \cite{Efimova2008,Dellis2012}. Moreover, radical pairs are never perfectly ordered in biological systems further reducing directional sensitivity \cite{Hill2010,Ilia2010,Lau2009,Lau2012}. So far, these models all assume that the hyperfine interactions themselves are constant.

However, proteins fluctuate on many time scales, with fluctuations generally being larger for slower time scales \cite{Frau2000}. Such protein movement affects quantum states of co-factors, as observed e.g.~in photosynthetic proteins \cite{Hofmann2010}. Thus, it is reasonable to expect fluctuations of quantum parameters (e.g.~hyperfine tensors) in the flavin co-factors over many time scales including times longer than radical pair reaction times. Moreover, one expects that these slow fluctuations are at least of the same size as fast fluctuations \cite{Hofmann2010}. Hyperfine interactions of flavin radicals, crucial for the functioning of a cryptochrome-based compass sensor \citep{Cinto2003,Lau2012}, were estimated to fluctuate on short time scales with an average magnitude of $0.6$ Gauss (G), due to the influence of the local protein environment \cite{Pauwels2010} and similarly sized fluctuations should occur over longer times as well. 

Here, we ask whether some nuclear spin environments promote signals that are more robust to slow hyperfine interaction fluctuations than others. Clearly, such environments are advantageous for a biological compass sensor because they prevent degradation of directional information without the need for additional, e.g.~neuronal, compensation mechanisms. Our study is extending and complementing studies investigating optimal nuclear spin environments for directional information \cite{Cai2012}, that have so far neglected the effects of such fluctuations expected to be present in realistic environments.

\section{\label{sec:theory}Theoretical model}
We study a photoinduced radical pair where the two electron spins interact only with the magnetic nuclei on their respective molecular moieties via hyperfine interactions and with an external magnetic field via the Zeeman interaction. Furthermore, the radicals are considered to form perfectly ordered crystals so as not to average out the anisotropy necessary for directional effects.

The singlet yield is calculated following the phenomenological density matrix approach, developed by the spin chemistry community \cite{Steiner1989} and successfully tested experimentally \cite{Book1}.

We consider a radical pair created in a spin-correlated singlet state, described by the spin density operator, $\hat{\rho}(0)=\hat{P}^{S}/Tr[\hat{P}^{S}]$, where $\hat{P}^S$ is the singlet projection operator. The coherent evolution of the pair is obtained from the Liouville-Von Neumann equation as $\hat{\rho}(t)=e^{-i\hat{H}t}\hat{\rho}(0)e^{+i\hat{H}t}$, with $\hat{H}$ denoting the spin Hamiltonian. The time dependent probability that the radical pair is in a singlet state is:
\begin{equation}
F_{S}(t)=Tr[\hat{P}^{S}\hat{\rho}(t)]=\dfrac{1}{M}\sum_{m=1}^{4M}\sum_{n=1}^{4M}|P^{S}_{mn}|^{2}\cos\left(\nu_{mn} t\right)
\end{equation}
where $M$ is the number of nuclear spin configurations and $P_{mn}^S=\langle m|\hat{P}^S |n\rangle$. Thus, the spin system oscillates via quantum coherences ($m\neq n$) between singlet and triplet states at frequencies $\nu_{mn}=\nu_{m}-\nu_{n}$ corresponding to the energy gap between pairs of eigenstates $|n\rangle$ and $|m\rangle$ of $\hat{H}$.
At the same time, singlet and triplet pairs recombine forming distinct singlet and triplet products.

Assuming a spin-independent first-order decay rate $k$ for the re-encounter probability distribution \cite{Kaptein1969,Rodgers2009}, we calculate the singlet yield, $\Phi$, as $\Phi= k \int_0^\infty\ F_{S}(t)~ e^{-kt} {\rm d}t$, which gives:
\begin{equation}
\Phi=\frac{1}{M}\sum_{m=1}^{4M}\sum_{n=1}^{4M}|P_{mn}^S|^{~2}
f(\nu_{mn})\label{eq:yield}
\end{equation}
with $f(\nu_{mn})=k^2/(k^2+\nu_{mn}^2)$ denoting a Lorentzian function.
A coherence $P_{mn}^S$ $(m\neq n)$ contributes to the biologically detectable signal $\Phi$ if its frequency has time to evolve before spin coherence is lost, i.e.~$|\nu_{mn}| \geq k$.

We study a class of radical pairs in which one radical is devoid of hyperfine interactions, whereas the other one interacts with the surrounding nuclei through axial hyperfine (HF) interactions.
This particular class has been suggested to be optimally sensitive and matches some of the functional characteristics of the avian magnetic compass observed in experiments \cite{Ritz2009,RitzRP}.

We consider two models of radical pair-based magnetic compass sensors: one is the simplest geomagnetic compass sensor comprising two electron spins $\hat{S}_{A}$ and $\hat{S}_{B}$ and one spin $1/2$-nucleus $\hat{I}$ coupled, without loss of generality, with spin A, whereas in the second model spin A interacts with two spin $1/2$-nuclei. 
The corresponding spin Hamiltonians $\hat{H}$ are:
\begin{eqnarray}\label{eq:H_tot}
\hat{H}(\theta)&=&\sum_{k}\hat{S}_{A}\cdot\mathbf{A}_{k}\cdot\hat{I}_{k}
+\omega(\cos\theta \hat{S}_{Az}+\sin\theta\hat{S}_{Ax})\nonumber\\
&+&\omega(\cos\theta \hat{S}_{Bz}+\sin\theta\hat{S}_{Bx})
\end{eqnarray}
with $k=1,2$ denoting the number of nuclei. We call these two models, respectively, one-HF ($k=1$) and two-HF ($k=2$) radical pair model. $\omega=\gamma_e B$ is the electron Larmor frequency in the applied magnetic field $B$, with $\gamma_e$ denoting the electron gyromagnetic ratio and $\mathbf{A}$ is an axial tensor given in its diagonal form by A=diag($\tilde{A}_{x},\tilde{A}_{y},\tilde{A}_{z}$). The following convention is used to parametrise the axial symmetry: $\tilde{A}_x=\tilde{A}_y=\tilde{a}-\tilde{a}\alpha$ and $\tilde{A}_z=\tilde{a}+2 \tilde{a}\alpha$, with $\tilde{a}$ being the strength and $\alpha$ the axiality of the hyperfine interaction. Finally, without loss of generality, $\theta$ is the polar angle between the external magnetic field and the radical pair system. 

For more realistic models, with more than two hyperfine interactions, the direction of the external magnetic field with respect to the radical pair system is defined by the azimuth and polar angles.
While the one-HF and two-HF models are special cases, they allow us to reduce the parameter space neglecting the azimuth angle. Thus, considering the dependence of the spin Hamiltonian on the polar angle $\theta$ in Equation (\ref{eq:H_tot}) we can calculate straightforwardly the signal $\Phi(\theta)$, from Eq.~\ref{eq:yield}, for these two models. We quantify angular sensitivity, $\Delta\Phi$, as the difference between the maximum $\Phi^{max}$ and minimum $\Phi^{min}$ of the singlet yield as a function of $\theta$, $\Delta\Phi=\Phi^{max} \left( \theta \right) -\Phi^{min} \left( \theta \right)$.

\section{\label{sec:Results}Results}
\subsection{\label{subsec:One-HF}One-HF radical pair model}
\begin{figure}[t!]
\includegraphics[width=0.9\linewidth]{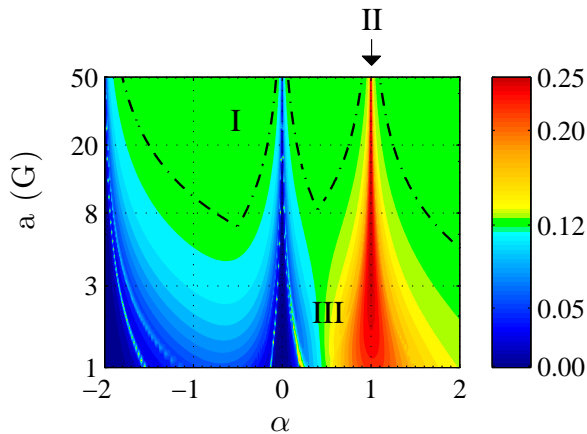}
\caption{(Color online) One-HF radical pair model: angular sensitivity $\Delta\Phi$ as a function of the strength, $a$ ($a=\tilde{a}/\gamma_e$), in Gauss (G), and axiality $\alpha$ of the hyperfine interaction. The rate constant is $k=10^{-1}$ MHz and the geomagnetic field is set to $0.5$ G. The angular sensitivity is calculated as $\Delta\Phi=\Phi^{max}(\theta)-\Phi^{min}(\theta)$. The dashed lines define the plateau region of angular sensitivity $\Delta\Phi=1/8$ (region I).}
\label{Fig:AS}
\end{figure}

For the one-HF radical pair model, the angular dependence of the singlet yield $\Phi(\theta)$ can be found analytically in the approximation of weak Zeeman interaction with respect to the hyperfine interaction, i.e.~for the electron Larmor frequency $\omega$ being much smaller than all zero-field quantum coherences,
and the rate constant $k$ slow enough to allow all quantum coherences to evolve, i.e.~$\lbrace\vert \tilde{A}_x \vert,~\vert \tilde{A}_x+\tilde{A}_z \vert /2,~ \vert \tilde{A}_x-\tilde{A}_z \vert / 2\rbrace \gg \omega \gg k$. In this approximation, which we summarize as $\tilde{a} \gg \omega \gg k$, two regions can be distinguished, as indicated in Fig.~\ref{Fig:AS}: region I, for $\tilde{a} \gg \omega \gg k$ with $\alpha\neq 0,-2,1$, where the signal $\Phi(\theta)\simeq \left(5+\cos 2\theta \right)/16$ \cite{Timmel2001, Ref_Probe} and the corresponding  angular sensitivity $\Delta\Phi \simeq 1/8$ are independent on the  hyperfine parameters; region II, for $\tilde{a} \gg \omega \gg k$ with $\alpha=1$, where the signal is $\Phi(\theta)\simeq \left(3+\cos 2\theta \right)/8$ and has an angular sensitivity of $\Delta\Phi \simeq 1/4$ \cite{Cai2012, Ref_Probe}. The signal from region II corresponds to a theoretical limit case, where the hyperfine tensor reduces to a vector. In region III the rate constant $k$ is slow enough to allow all quantum coherences to evolve, and the electron Larmor frequency is almost of the same order of magnitude as  all zero-field quantum coherences, i.e.~$\lbrace\vert \tilde{A}_x \vert,~\vert \tilde{A}_x+\tilde{A}_z \vert /2,~ \vert \tilde{A}_x-\tilde{A}_z \vert/ 2\rbrace ~\gtrsim~ \omega \gg k $. We summarize  region III as $\tilde{a}~\gtrsim~ \omega \gg k$ with $\alpha\neq 0,-2$. In region III the analytical expression of the angular dependence of the singlet yield is more complicated and depends strongly on the hyperfine parameters.

We numerically evaluate the angular sensitivity for hyperfine coupling strengths $a$ ($a=\tilde{a}/\gamma_e$) between $1$ and $50$ G, i.e.~a range of values typically encountered in organic radicals, and all possible values for the axiality $\alpha$.

Figure \ref{Fig:AS} shows the angular sensitivity $\Delta\Phi$ as a function of strength $a$ and axiality $\alpha$ of the hyperfine interaction. The rate constant is $k=10^{-1}$ MHz and geomagnetic field strength $B$ is set to $0.5$ G. Numerical calculations are in agreement with the analytical ones and give more information on the signals generated by the one-HF radical pair model. We note that region I, represented in Fig.~\ref{Fig:AS} by the green color and defined by the dashed lines, is a wide plateau region of angular sensitivity $\Delta\Phi=1/8$ that occurs for strengths $a\gg 10$ G and axialities $\alpha\neq 0,-2,1$. We also note that region II, with an angular sensitivity of $\Delta\Phi=1/4$, occurs for hyperfine strengths $a \gg 5$ G and axiality $\alpha=1$ in agreement with the results in \cite{Cai2012} and represents the highest angular sensitivity reachable within the hyperfine interactions range considered here. We will see that such signal, although showing the highest angular sensitivity, is strongly affected by fluctuations of hyperfine tensor. Furthermore, from Fig.~\ref{Fig:AS} we note region III roughly corresponding to hyperfine strengths $a < 10$ G and axiality $\alpha\neq 0,-2$. This region contains a wide range of angular sensitivity values because the signals depend strongly on the HF parameters, especially on the axiality. Greater axialities, $\alpha> 0$, enhance angular sensitivity, whereas axialities with $\alpha < 0$ reduce angular sensitivity. We perform the same calculations for different rate constants, i.e.~$k=10^{-2},10^{-3},10^{-4}$ MHz and find qualitatively identical results. So far, signals and corresponding directional sensitivities are calculated neglecting hyperfine fluctuations.

It was found that fluctuations of hyperfine tensors in flavin radicals \cite{Pauwels2010} follow roughly Gaussian distributions, with an estimate of $0.6$ G on average for the standard deviations. Taking this estimation, to simulate hyperfine tensor fluctuations, for each strength $a$ and axiality $\alpha$ we draw a set of hyperfine tensor components $\lbrace(A_{xn},A_{zn})\rbrace$, with $n=1,3000$, from two Gaussian distributions with respective means $A_{x}=a-a\alpha$ and $A_{z}=a+2a\alpha$ and identical standard deviation of $0.6$ G. 
We consider such fluctuations on a timescale longer than the radical pair lifetime (fluctuations much faster than quantum coherences are averaged out), which we choose to be $10~ \mu s$. Thus, for each set of hyperfine tensor fluctuations we calculate the corresponding set of $3000$ angular dependences of singlet yields $\lbrace\Phi(\theta)_n\rbrace$, which represents the directional signal at different times $t_n$, with $\theta$ ranging from $0^{\circ}$ to $180^{\circ}$. We then calculate the angular dependence of the mean singlet yield $\bar\Phi(\theta)$ and corresponding standard deviation $\sigma(\theta)$ by averaging over the $3000$ samples for each angle $\theta$. Finally, we calculate the mean standard deviation over the angles as $\bar{\sigma}=((\sum_{\theta_i=0}^{180}\sigma(\theta_i)/i)$, which we use as a measure of robustness of the directional signal.
\begin{figure}[tb!]
\centering
\subfigure[]{
\includegraphics[width=0.67\linewidth]{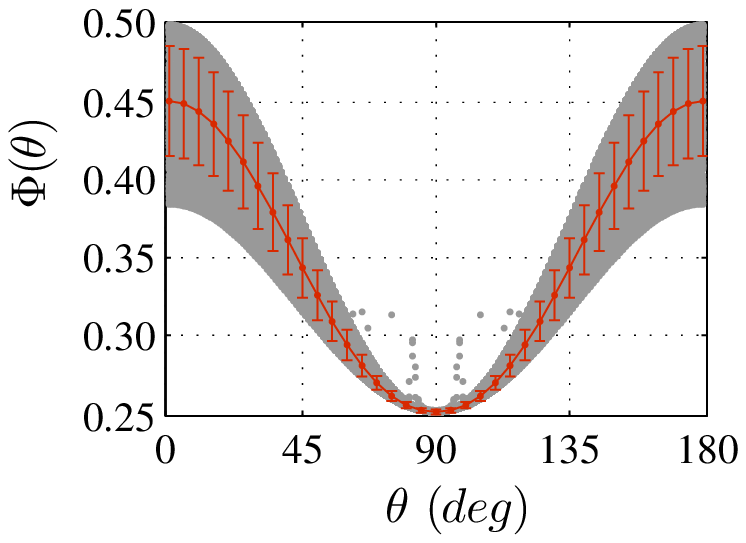}}
\subfigure[]{
\includegraphics[width=0.67\linewidth]{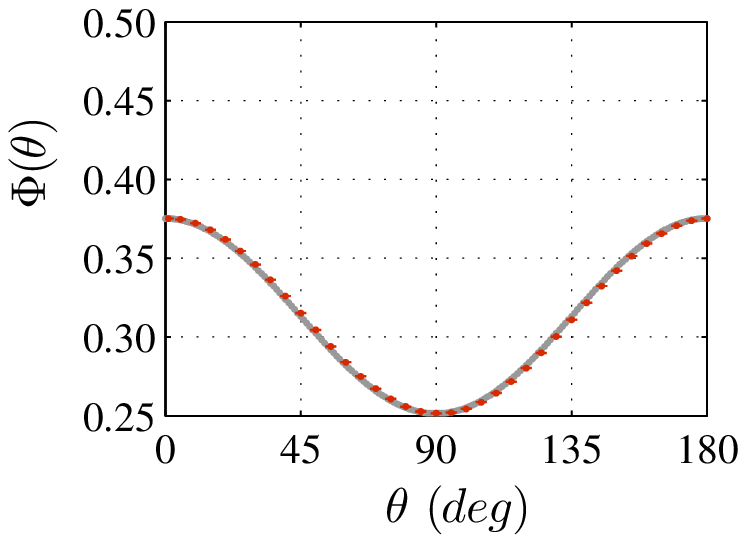}}
\subfigure[]{
\includegraphics[width=0.48\linewidth]{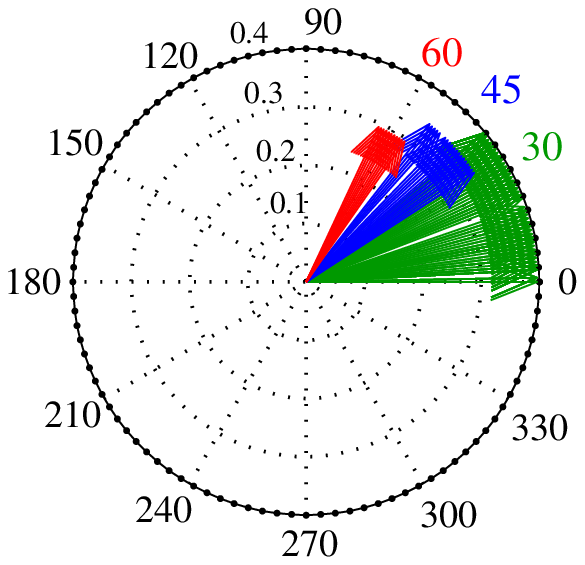}}
\subfigure[]{
\includegraphics[width=0.48\linewidth]{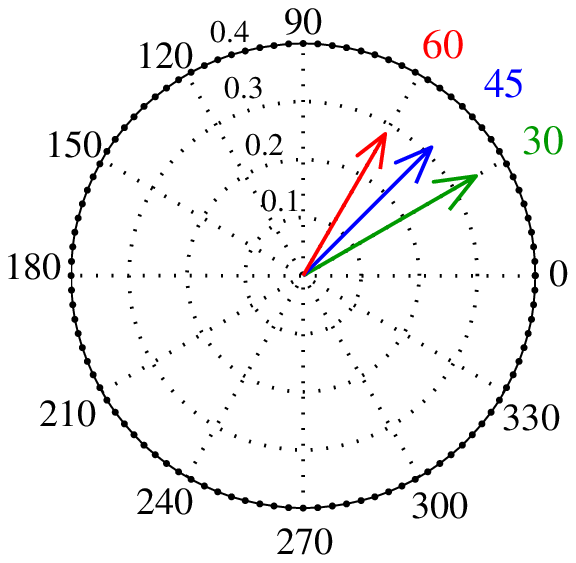}}
\caption{(Color online) One-HF interaction model: examples of (a) a non-robust signal (robustness $\bar{\sigma}= 10^{-2}$) from region II, with hyperfine strength $a=8$ G and axiality $\alpha=1$ and (b) a robust signal ($\bar{\sigma}= 10^{-4}$) form region I, with $a=8$ G and $\alpha=-0.6$, under a set of $3000$ Gaussian samples of hyperfine tensor fluctuations slower than the radical pair lifetime, chosen to be $10~\mu s$. The gray area represents the corresponding set of $3000$ angular dependences of singlet yields $\lbrace\Phi(\theta)_{n}\rbrace$, which represents the directional signal at different times $t_{n}$, $n=1,3000$. The red line indicates the angular dependence of the mean singlet yield $\bar\Phi(\theta)$. The bars correspond to the standard deviation $\sigma(\theta)$ for each angle $\theta$. The robustness of the signal $\bar{\sigma}$ is the mean standard deviation over the angles. (c) Non-reliable directional information of a non-robust signal. A specific singlet yield value (length of the arrow) corresponds to a different angle at different times. The values are those used in (a). (d) Reliable directional information of a robust signal. A specific singlet yield value signals the same angle at different times. The values are those used in (b). One arrow depicts a value of the singlet yield at an angle $\theta$ and at time $t_n$. Each color corresponds to a different singlet yield value.}
\label{Fig:examples}
\end{figure}

Figure \ref{Fig:examples} shows an example of (a) a non-robust signal and (b) a robust signal, with the red line indicating the angular dependence of the mean singlet yield. The bars correspond to the standard deviation for each angle. The gray area represents the angular dependence of the singlet singlet yield at different times. For the non-robust signal in Fig.~\ref{Fig:examples} (a) the mean standard deviation is $\bar{\sigma}=10^{-2}$, which means a variation of $1\%$ in the angular dependence of the singlet yield. This implies that a specific singlet yield value corresponds to a different angle at different times, as is depicted in Fig.~\ref{Fig:examples} (c), which reports some of the values used in Fig.~\ref{Fig:examples} (a), with the length of the arrow representing a singlet yield value and the direction the corresponding angle. 
For the robust signal in Fig.~\ref{Fig:examples} (b) the mean standard deviation is $\bar{\sigma}=10^{-4}$, which corresponds to a variation of $0.01\%$ in the angular dependence of the singlet yield. 
A robust signal has reliable directional information, since a specific singlet yield signals the same angle at different times, as depicted in Fig.~\ref{Fig:examples} (d) which reports some values of Fig.~\ref{Fig:examples} (b). 

\begin{figure}[b!]
\subfigure[]{
\includegraphics[width=0.87\linewidth]{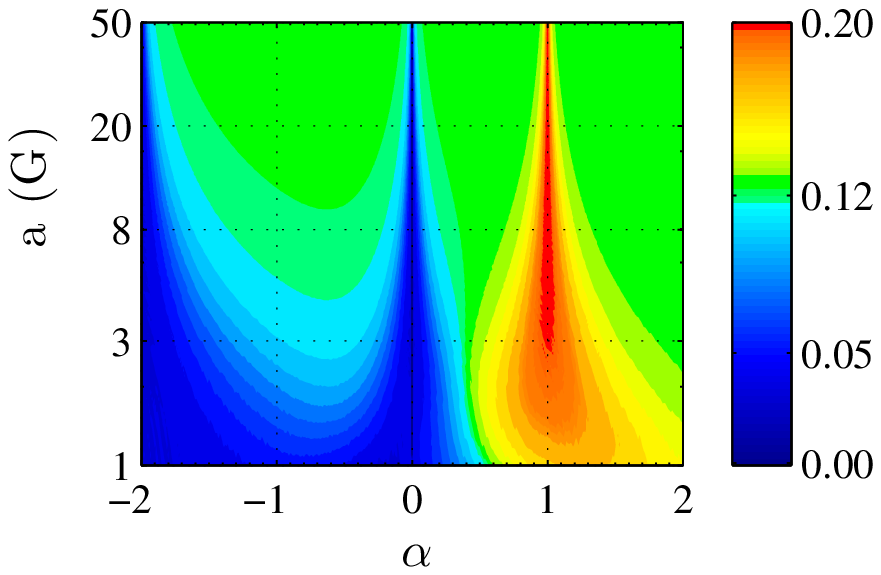}}
\subfigure[]{
\includegraphics[width=0.87\linewidth]{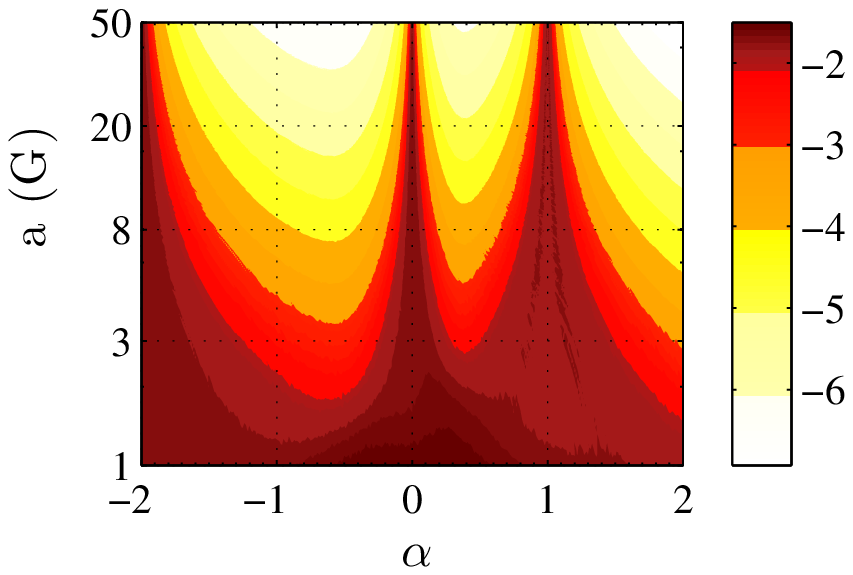}}
\caption{(Color online) One-HF radical pair model under hyperfine tensor fluctuations slower than the lifetime of the radical pairs, chosen to be $10~\mu s$. (a) Mean angular sensitivity $\Delta\bar{\Phi}$ as a function of the hyperfine strength $a$ and axiality $\alpha$. (b) Robustness $\bar{\sigma}$, in logarithm units, as a function of $a$ and $\alpha$. The geomagnetic field is set to $0.5$ G.}
\label{Fig:ASmstdYm}
\end{figure}

Figure \ref{Fig:ASmstdYm} (a) reports the mean angular sensitivity $\Delta\bar{\Phi}$, evaluated as $\Delta\bar{\Phi}=\bar{\Phi}(\theta)^{max}-\bar{\Phi}(\theta)^{min}$, as a function of strength $a$ and axiality $\alpha$ of the hyperfine interaction.
Comparing Fig.~\ref{Fig:AS} and Fig.~\ref{Fig:ASmstdYm} (a) we note that, under fluctuations, the mean angular sensitivity $\Delta\bar{\Phi}$ remains quantitatively similar to $\Delta\Phi$ without fluctuations. Indeed, we can still discriminate the three regions described for the static case. However, we note that fluctuations of hyperfine tensor components give rise to a lower mean angular sensitivity, in general, for hyperfine axiality $\alpha \approxeq 1$.
Furthermore, Fig.~\ref{Fig:ASmstdYm} (b) depicting the robustness $\bar{\sigma}$ in logarithm units as a function of $a$ and $\alpha$, shows that for the one-HF radical pair model, region II and region III are not robust with corresponding signals highly affected by fluctuations. Only region I, the plateau region for the static case (cf. Fig.~\ref{Fig:AS}) is robust against fluctuations. We find that such a robustness persists under fluctuations of $1$ G standard deviation, which is the largest standard deviation found for the hyperfine tensor fluctuations in the flavin radicals. For the one-HF model, the wide plateau region guarantees robustness to a wide range of magnitude of hyperfine tensor fluctuations.

We perform the same calculations for slower rate constants, i.e.~$k=10^{-2},10^{-3},10^{-4}$ MHz, and find the same results as for $k=10^{-1}$ MHz, with region I being still the only robust region. Analytical evaluation shows that this robustness is a general feature for large hyperfine interaction and slow $k$.
We calculate a more general expression of the singlet yield as a function of $\theta$, $\Phi(\theta)$, i.e.~in the approximation of weak Zeeman interaction with respect to the hyperfine interaction and a rate constant $k$ slow enough to allow all zero-field quantum coherences to evolve, which we summarize as $ \tilde{a} \gg \omega$  and $\tilde{a} \gg k$, with $\alpha\neq 0,-2,1$. In this approximation the singlet yield is:
\begin{eqnarray}\label{Eq:yield_approx}
\Phi(\theta)&\simeq &\frac{3}{8}-\left[\frac{1}{16}-\frac{1}{16} f(\omega)\right]
\left(1-\cos 2\vartheta\right)
\end{eqnarray}
which gives a more general expression for the angular sensitivity as:
\begin{eqnarray}\label{Eq:ang_sens_approx}
\Delta\Phi\simeq\frac{1}{8}-\frac{1}{8}f\left(\omega\right)=\frac{1}{8}-\frac{1}{8}\dfrac{k^{2}}{(k^{2}+\omega{^2})}
\end{eqnarray}
Considering the ratio $k/\omega$ in Eq.~\ref{Eq:ang_sens_approx}, when the external field also has time to mix the states, i.e.~$\tilde{a} \gg \omega \gg k$ with $\alpha\neq 0,-2,1$, the signal becomes independent on the rate constant, i.e.~$\Phi(\theta)\simeq \left(5+\cos 2\theta \right)/16$, and the angular sensitivity becomes $\Delta\Phi\simeq1/8$.

\subsection{\label{subsesec:two-HF}Two-HF radical pair model}
The one-HF radical pair model has been extensively studied as a proof of principle in the literature \cite{Timmel2001,Kasz2012,Cai2012,Kominis2009,Dellis2012,Vedral2011}.

Here, we consider a slightly more complex situation in which there are two axial hyperfine interactions in one radical, whereas the other radical is devoid of hyperfine interactions. We find that when both radicals have one hyperfine interaction each the angular sensitivity is substantially suppressed \cite{2HFcalc}.

We randomly generate $10^5$ samples of two-HF radical pairs with hyperfine strengths, $a_{1}$, $a_{2}$, ranging from $1$ to $50$ G and axialities, $\alpha_{1}$, $\alpha_{2}$, ranging from $-2$ to $2$. The angle $\psi$ between the two hyperfine interactions is drawn randomly from $0^\circ$ to $90^\circ$. We use a rate constant of $k=10^{-1}$ MHz and a geomagnetic field strength of $0.5$ G, as in the one-HF radical pair model.
We select the samples having the highest angular sensitivity of $0.25$ and we search for hyperfine patterns. We find two patterns which lead to an optimal angular sensitivity. The first pattern consists of two hyperfine strengths whose ratio is roughly within one order of magnitude, axialities both close or equal to $1$ and collinear axes. The second pattern, already supposed to be optimal \cite{Cai2012}, consists of one dominant hyperfine interaction with axiality close or equal to $1$ and the second hyperfine interaction being a small perturbation. We find the same patterns for slower rate constants.
\begin{figure}[h!]
\centering
\subfigure []{
\includegraphics[width=0.8\linewidth]{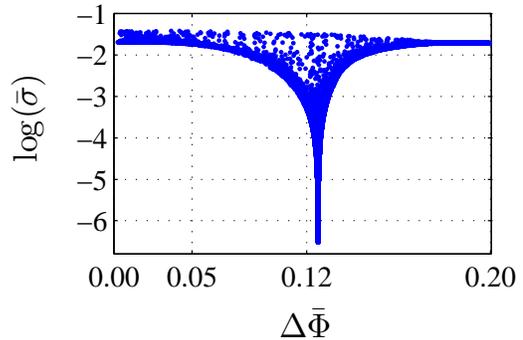}} 
\subfigure[]{
\includegraphics[width=0.8\linewidth]{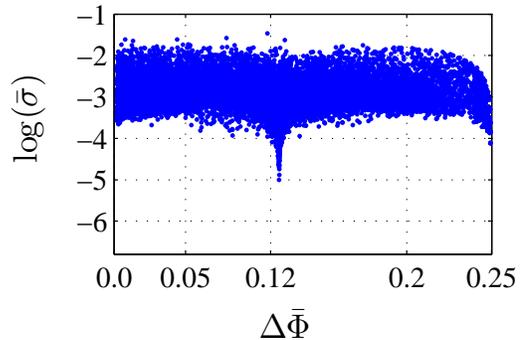}}
\caption{(Color online) Robustness $\bar{\sigma}$, in logarithm units, vs the mean angular sensitivity $\Delta\bar{\Phi}$ for (a) one-HF radical pair model and (b) two-HF radical pair model, under fluctuations of hyperfine tensor components slower than the radical pair lifetime, chosen to be $10~ \mu s$. The geomagnetic field is set to $0.5$ G. In (a) $\Delta\bar{\Phi}$ and $\bar{\sigma}$ are the values used in Fig.~\ref{Fig:ASmstdYm} (a) and Fig.~\ref{Fig:ASmstdYm} (b). In (b) $\Delta\bar{\Phi}$ and $\bar{\sigma}$ are calculated over $10^5$ samples of two-HF model with random strengths and axialities as described in the text.}
\label{Fig:Hists}
\end{figure}

To simulate slow hyperfine tensor fluctuations, for each of the $10^5$ two-HF radical pair samples ($a_1,a_2,\alpha_1,\alpha_2,\psi$)  we draw $3000$ Gaussian random samples for each hyperfine tensor components, with respective means of $A_{x1}, A_{z1}, A_{x2}, A_{z2}$, and identical standard deviation of $0.6$ G. For each of the $10^5$ two-HF radical pair samples, we calculate the angular dependence of the mean singlet yield $\bar{\Phi}(\theta)$ and corresponding mean angular sensitivity $\Delta\bar{\Phi}=\bar{\Phi}(\theta)^{max}-\bar{\Phi}(\theta)^{min}$, using a rate constant $k=10^{-1}$ MHz. The robustness of signals is measured by the mean standard deviation $\bar\sigma$ as with the one-HF model. Figure \ref{Fig:Hists} $(b)$ reports the robustness $\bar\sigma$, in logarithm units, as a function of the mean angular sensitivity $\Delta\bar{\Phi}$ of the $10^5$ two-HF radical pair samples. Figure \ref{Fig:Hists} $(a)$ reports the same parameters for the one-HF radical pair model, as used in Fig.~\ref{Fig:ASmstdYm}. Comparing these two models, from Fig.~\ref{Fig:Hists} (a) and (b), we note that under the same magnitude of fluctuations, i.e.~$0.6$ G standard deviation, the most robust signals have mean angular sensitivity of $\Delta\bar{\Phi}=0.125$, for both models, with a stronger robustness for the one-HF model. Furthermore, in the two-HF model the robustness increases over a wide range of angular sensitivities compared to the one-HF model. Finally, only the two-HF model generates some signals with an optimal mean angular sensitivity of $0.25$ and robustness of $\log(\bar{\sigma}) \leq -4$, i.e.~optimal signals that have a variation of $0.01\%$ in the angular dependence of the singlet yield under fluctuations.

For the two-HF model, such a robustness of signals with mean angular sensitivities of $0.125$ and $0.25$ persists under fluctuations of $1$ G standard deviation. Optimal signals with a robustness of $\log(\bar{\sigma}) \leq -4$, are given by the first pattern mentioned above. The second pattern of optimal signals, i.e.~with one dominant hyperfine interaction, shows a weaker robustness. We find identical results for slower rate constants.

\section{\label{sec:end}Conclusions} 
We have investigated the effect of slow fluctuations in the protein
environment and, hence, hyperfine interactions, on the signals arising
from a radical-pair based compass sensor. Evaluating these effects on two
radical-pair models with a wide parameter range reveals that the nuclear
spin environment has a significant influence on the size of perturbative
effects of such fluctuations. In fact, some nuclear spin environments can
make a radical-pair based compass sensor virtually immune to the influence
of slow fluctuations.

Such robustness is a design feature separate from optimality. An optimal
compass sensor can be found when the directional sensitivity is maximized (maximum angular sensitivity), whereas a robust sensor is one where the directional sensitivity remains
unaffected by fluctuations regardless of the magnitude of directional
sensitivity. The question arises whether one can maximize both optimality
and robustness at the same time. In the one-HF radical pair model, the
optimal choice of parameters (maximal anisotropy of the hyperfine interaction), resulting in
an angular sensitivity of $0.25$, does not turn out to be robust. However, the second best
choice, resulting in an angular sensitivity of $0.125$ is robust, whereas other parameters
with even lower sensitivities reduce robustness as well. This indicates
that there is not a simple trade-off between optimality and robustness
with one increasing while the other decreases.

The interplay between optimality and robustness becomes more complex when
multiple hyperfine interactions are involved. Perhaps the most noteworthy
observation from our results is that simply adding another hyperfine
interaction can increase robustness over a wide range of parameters,
while leaving the sensitivity largely unaffected, and that there are 'sweet
spots' where high optimality and robustness can be achieved. Further studies will
need to confirm whether this trend continues for larger numbers of
hyperfine interactions. The goal of this manuscript is to raise the issue
of robustness in addition to optimality in the discussion of
quantum-based magnetic compass sensors.

\begin{acknowledgements}
This work was supported by a grant from DARPA. We would like to tank E. Pauwels for helpful comments and making data available to us.
\end{acknowledgements}


\begin{thebibliography}{43}
\bibitem{2HFcalc}We randomly generated $10^5$ samples of two-HF radical pair model with hyperfine strengths ranging from $1$ to $50$ Gauss and axiality ranging from $-2$ to $2$. We compute the yield of each $10^5$ samples for different rate constants $k = 10^0, 10^{-1}, 10^{-2}, 10^{-3}$ MHz, and for two cases: case 1HF-1HF, with one hyperfine interaction on each radical and case 2HF-0, with two hyperfine interactions in the same radical. The results show that the 2HF-0 case has,
generally, a larger angular sensitivity than 1HF-1HF case for all rate constants considered. Samples in which the angular sensitivity for 1HF-1HF is larger than 2HF-0 are noisy signals. Magnetic field effects reach saturation for a rate constant slower than $10^{-1}$ MHz for both cases.
\end{thebibliography}
\end{document}